\documentclass{article}

\usepackage{arxiv}

\usepackage[utf8]{inputenc} % allow utf-8 input
\usepackage[T1]{fontenc}    % use 8-bit T1 fonts
\usepackage{hyperref}       % hyperlinks
\usepackage{url}            % simple URL typesetting
\usepackage{booktabs}       % professional-quality tables
\usepackage{amsfonts}       % blackboard math symbols
\usepackage{nicefrac}       % compact symbols for 1/2, etc.
\usepackage{microtype}      % microtypography
\usepackage{lipsum}		% Can be removed after putting your text content
\usepackage{graphicx}
\usepackage{natbib}
\usepackage{doi}
\usepackage{threeparttable}
\newcommand{\figlabel}{Figure}
\usepackage{soul}
\usepackage{enumitem}
\usepackage{multirow}

\title{SciEv: Finding Scientific Evidence Papers for Scientific News}

\author{Md Reshad Ul Hoque\\
	Electrical \& Computer Engineering\\
	Old Dominion University\\
	Norfolk, VA, USA \\
	\texttt{mhoqu001@odu.edu} \\
	\And 
	Jiang Li\\
	Electrical \& Computer Engineering\\
	Old Dominion University\\
	Norfolk, VA, USA \\
	\texttt{jli@odu.edu} \\
	\And
	Jian Wu \\
	Computer Science\\
	Old Dominion University\\
	Norfolk, VA, USA \\
	\texttt{jwu@cs.odu.edu} \\
}

\renewcommand{\shorttitle}{SciEv: Finding Scientific Evidence Papers for Scientific News}

\hypersetup{
pdftitle={SciEv: Finding Scientific Evidence Papers for Scientific News},
pdfsubject={cs.AI,cs.DL},
pdfauthor={Md Reshad Ul Hoque, Jian Wu},
pdfkeywords={Scholarly papers, Fake news,Web api, Deep learning, Domain knowledge entity, Embedding, Transformer},
}

\begin{document}
\maketitle

\begin{abstract}
	In the past decade, many scientific news media that report scientific breakthroughs and discoveries emerged, bringing science and technology closer to the general public. However, not all scientific news article cites proper sources, such as original scientific papers. A portion of scientific news articles contain misinterpreted, exaggerated, or distorted information that deviates from facts asserted in the original papers. Manually identifying proper citations is laborious and costly. Therefore, it is necessary to automatically search for pertinent scientific papers that could be used as evidence for a given a piece of scientific news. We propose a system called SciEv that searches for scientific evidence papers given a scientific news article. The system employs a 2-stage query paradigm with the first stage retrieving candidate papers and the second stage reranking them. The key feature of SciEv is it uses domain knowledge entities (DKEs) to find candidates in the first stage, which proved to be more effective than regular keyphrases. In the reranking stage, we explore different document representations for news articles and candidate papers. To evaluate our system, we compiled a pilot dataset consisting of 100 manually curated (news,paper) pairs from ScienceAlert and similar websites. To our best knowledge, this is the first dataset of this kind. Our experiments indicate that the transformer model performs the best for DKE extraction. The system achieves a P@1=50\%, P@5=71\%, and P@10=74\% when it uses a TFIDF-based text representation. The transformer-based re-ranker achieves a comparable performance but costs twice as much time. We will collect more data and test the system for user experience. 
\end{abstract}

% keywords can be removed

\section{Introduction}
Scientific news articles report scientific and technological progresses, breakthroughs, discoveries, and innovations. Scientific news is important media to disseminate scientific knowledge to the general public. Different from other types of news, such as political news, scientific news is believed to be written in the way such that it faithfully presents facts from the source. However, this may not always be the case. For example, exaggerations in popular scientific writing could misinform the public and even researchers \cite{west2021misinformation}. Since the start of Covid-19, public support for policies to control the spread of this virus is being undercut by misinformation, leading to the World Health Organization's ``infodemic'' declaration \cite{zarocostas2020infodemic}. Scientific misinformation could be spread orally or over social media, causing social problems and further damage the credibility of scientific news. %One example is the misunderstanding that mosquitoes transmit coronavirus, which was revealed wrong by a recent study \cite{huang2020mosquitoes}.  
Effectively curbing scientific misinformation is crucial to avoid these problems. 
%One method is to automatically find scientific papers as references for a given news article. 

%Scientific articles, containing scientists' understanding of domain knowledge, analytical and experimental results, and new discoveries are traditionally read by researchers whose research is relevant to the particular domains the articles fall into. This is to a large extent due to the language used in scientific articles, including the appearance of jargon and terminologies as well as the complicated methods and datasets that are difficult to interpret without enough professional training. As a result, scientific research is usually believed to be a field far away from the general public. Therefore, many scientific breakthroughs remain not well-known until they are ``translated'' into more readable text via scientific news. This can cause a gap between people's common beliefs, which may be prevalent but not up to date with the recent scientific research outcome. One example is the misunderstanding that mosquitoes transmit coronavirus, which was revealed wrong by a recent study \cite{huang2020mosquitoes}. 

In the past years, many websites that report scientific results in form of news articles have been launched. Examples of these websites include \url{ScienceAlert.com} and \url{ScienceDaily.com}. Many articles published on these websites contain references to source papers. These references provide scientific evidence to the news content and boost the credibility of scientific news articles. However, the majority of these hyperlinks were added manually and scientific news articles on other websites and social media do not always contain references to source papers. A fraction of these articles may contain misinterpreted, exaggerated, or distorted information that deviates from the truth revealed by original papers \cite{augenstein2021determining}. Finding the original papers may not be always possible. However, it is possible to develop computational models to automatically find pertinent scientific papers to given scientific news. 

This can be generally classified as a citation recommendation problem, the goal of which is to find an article that should be cited given a piece of context \cite{farber2020citation}. 
%A specific type of this problem is news citation recommendation in which the cited article is another news article. 
%Citation recommendations can be dealt with using supervised or unsupervised methods. 
In proposed solutions, a function is trained to map a citation context $z_{ij}$ and the document it belongs to ($d_i$) to a reference (aka the cited document $r_m$). In a supervised method, a classifier is trained to incorporate global (document or cross-document) and local (context) features, e.g., \cite{he2010context}. In an unsupervised method, a re-ranking model is applied which assigns probabilistic scores to a list of candidate documents, e.g., \cite{peng2016news}. 

Existing citation recommender systems focused on problems in which both original or recommended documents are either news articles \cite{peng2016news} or scientific articles \cite{he2010context}.
%\hl{<--I did not get the meaning of this sentence}. 
In our problem, the original article is a news article and the recommended articles are  scientific papers. This problem is more challenging for the following reasons. First, there is usually a gap between vocabularies used in news articles and scientific papers because reporters usually need to paraphrase scientific papers into more readable text for non-domain experts. The second challenge is data sparsity. Unlike dense citation networks \cite{kataria2010aaai}, there are much fewer citation relations between news and scientific articles, making it challenging to apply graph embedding methods. Lastly, to our best knowledge, there are few open-access datasets that can be used for training and evaluation. Recently, a dataset containing $\sim800$ peer-reviewed articles were compiled with mentions of scientific research in popular media but the original news articles were not included \cite{anderson2020case}. To overcome this challenge, we compiled a pilot dataset consisting of 100 pairs of scientific news articles and their associated research papers. 
%\hl{[<-- this is an important contribution that should be emphasized somewhere early in the paper, i.e., Abstract. ]}

In this paper, we proposed a system called SciEv as an intermediate step towards evaluating the credibility of scientific news. SciEv automatically finds scientific evidence papers for a given scientific news article in two stages. In the first stage, the system sends queries 
%\hl{why plural here?} 
to a digital library API and retrieve candidate papers. The second stage reranks these candidates based on their semantic similarities to the news article. To overcome the vocabulary gap challenge aforementioned, instead of querying general keyphrases, the system queries domain knowledge entities (DKEs) extracted from news articles \cite{hoque2019searching}. We queried DKEs instead of regular keywords because many DKEs appear both in news articles and corresponding scientific papers. This motivates us to use DKEs to search for scientific papers given a scientific news article. 
%\hl{<--difficult to follow, please double check}. 

\begin{figure}
    \centering
    \includegraphics[width=0.7\textwidth]{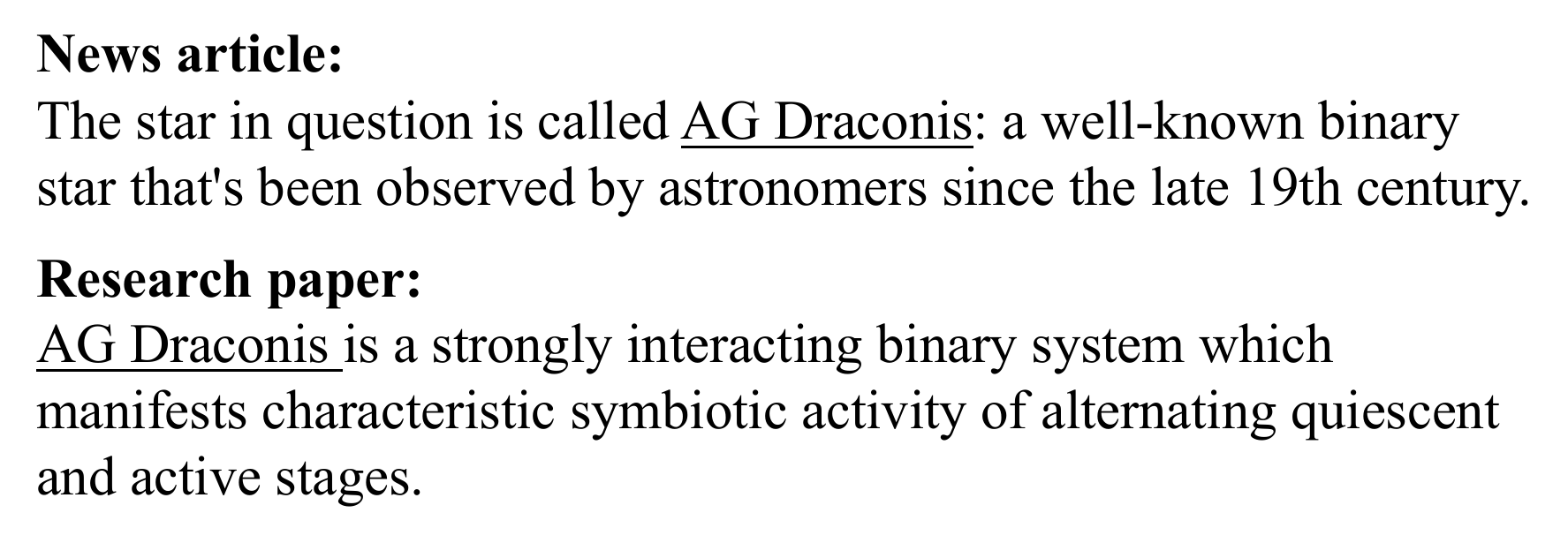}
    \caption{An example of a DKE (underlined) appearing in both a news article and a research paper.}
    \label{fig:example}
\end{figure}

\paragraph{\textbf{DKEs}} We define DKEs as noun phrases that deliver domain knowledge. DKEs are different from general named entities and keyphrases \cite{he2018keyphrase} because they are predominantly used in domain specific articles and usually need more or less domain knowledge to understand. In Figure~\ref{fig:example}, we highlight DKEs that appear in both a piece of scientific news and the corresponding paper. %Key phrases provide a top-level description of a paper. In general, an article may contain more DKEs than Key-phrases.On the other hand, name entities such as people, organizations, and locations, etc. generally provide general information rather than domain knowledge which can be extracted by Name Entity Recognition (NER) tools.
%\begin{quote}\small
%\ul{Dual-energy computed tomography (DECT)} has been widely used in many applications that need \ul{material decomposition}. The concept of \ul{dual-energy CT} was initially described in 1973 by {\color{red}Godfrey Hounsfield}.”. \end{quote}

%In our preliminary work, we proposed a method to extract DKEs trained on a single domain dataset and retrieve pertinent papers given a piece of scientific news using term frequency-inverse document frequency (TFIDF) representations. 
%\hl{[Check the tense in this paragraph, sometimes Present sometimes Past] }
%In this paper, we proposed an effective and robust system. The first stage used a transformer-based model trained on a manually labeled multi-disciplinary dataset for DKE extraction and improved the DKE extraction performance to F1$>92\%$. In the second stage, document embedding was employed to represent news articles and papers. We compared the efficacy and efficiency of different state-of-the-art language models on this task and evaluated the system with different settings. In the best setting, the P@10 metric was improved by 25\% and the overall runtime was reduced by 43\%, compared with the state-of-the-art system. 

%\begin{enumerate}
%    \item We propose a 2-stage retrieval system including a neural ranking model that re-ranks candidate scientific articles obtained by querying using a frequency-based ranker. The model is integrated into a system called SciPEP that recommends pertinent scientific papers given a scientific news article. 
%    \item We propose a transformer-based model that improves the performance of the DKE extraction.
%    \item We contribute a benchmark dataset that contains around 100 pairs of scientific news articles mostly from ScienceAlert.com and their associated research papers. 
%\end{enumerate}

\section{Related Work}
The exponential growth of scientific papers each year poses a great challenge for news editors and researchers to find the most pertinent citations \cite{bornmann2015growth}. Citation recommendations can be broadly classified into three categories depending on the source articles and the articles to be cited. In the first type, a news article cites another news article (news$\rightarrow$news). In the second type (paper$\rightarrow$paper), a scientific paper cites another scientific paper. In the third type (news$\rightarrow$paper), a news article cites a scientific paper. Citing a news article in a research paper is relatively uncommon and beyond the scope of our study. 

The collaborative filtering (CF) method has been widely used for news recommender systems since it was proposed \cite{melville2002aaai}. This method requires building the document and the user profiles. However, the reading history of news articles is usually unknown, making it difficult to build user profiles. 

Many citation recommender systems were proposed using different text representation models. Early work used Synset Frequency-Inverse Document Frequency (SF-IDF) to represent the news \cite{capelle2012semantics}. SF-IDF was similar to TFIDF except that it used WordNet synonym sets to expand the semantic representation of a given word. Other types of work leverage word embeddings. For example, Peng et al. (2016) developed a news citation recommendation system using a word-embedding based re-ranking and grounded entities (i.e., explicit semantics) \cite{peng2016news}. Okura et al. (2017) proposed a model called embedding-based news recommendation (EBNR) using the denoising autoencoders variant for news representations \cite{kumar2017deep}. Wang et al. (2018) utilized news titles and entities to represent the news via a knowledge-aware convolutional neural network (CNN) \cite{wang2018dkn}. Saskr et al. (2019) developed a news recommendation system that combines news titles and bodies using average embeddings \cite{chu2019next}.

Depending on the type of input, citation recommendation systems can be classified into local and global citation recommendations. Local citation recommendation systems are based on text snippets, such as a sentence or even several words \cite{huang2015neural}. Global citation recommendation systems make recommendations based on the full text or the abstract of a document \cite{kataria2010utilizing}.

We propose the Scientific Evidence paper retrieval system called SciEv, which can be classified as a news$\rightarrow$paper recommender system based on global text. To our best knowledge, few systems have been proposed with the same functionality \cite{hoque2019searching}. Recently, many pre-trained language models were proposed (e.g., BERT \cite{devlin2018bert}) and have shown efficacy in retrieval tasks by representing text with distributed vectors, e.g., \cite{zhang2020covidex}. This motivates us to compare these language models in the news$\rightarrow$paper recommender system. 
%In addition, the entity extractor in \cite{hoque2019searching} was trained only on biomedical domain using a sequence tagging model. We improve the entity extraction by training it with a transformer model on a multi-domain dataset. Such a dataset is necessary to support DKE extraction from scientific news in multiple domains. 

%In a traditional citation recommendation system, the recommended articles are usually used for supporting the context or the source article. Our system, instead, searches for pertinent scientific papers that could support or refute the given news article. The refute articles will play a more important role for news articles containing misinformation that is contrary to or inconsistent with scientific findings. 
%We propose using DKEs as queries to search for highly relevant papers. Our system can be built either on a local digital library search platform or an online search API. We propose a 2-stage information retrieval model that combines a frequency-based and a neural network-based ranker.  We compare recently proposed pre-trained language models that best represent news articles and scientific papers.  

\begin{figure}
  \centering
  \includegraphics[width=\textwidth]{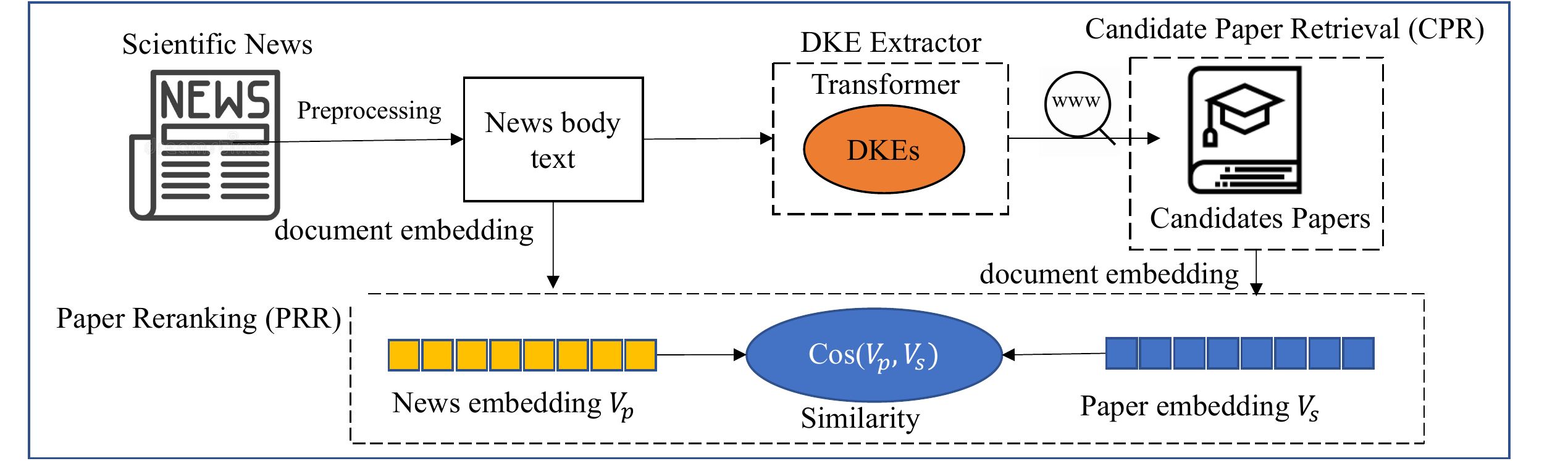}
  \caption{The architecture of the SciEv system.}
  %\Description{Enjoying the baseball game from the third-base seats. Ichiro Suzuki preparing to bat.}
  \label{fig:arch}
\end{figure}
\section{System Overview}
The architecture of the SciEv system adopted a two-stage retrieval model proposed, e.g.,  \cite{zhang2020covidex}. The architecture of the system contains four modules (Figure~\ref{fig:arch}) as described below.

\paragraph{\textbf{Preprocessing}} The input of this module is an HTML page of a scientific news article, downloaded from the Web. Only textual content is retained for further analysis. In the experiments below, we used a preprocessor that parses web pages downloaded from ScienceAlert.com. The parser could easily be customized to parse news body text from other websites. The text was then cleansed so square brackets, extra spaces, special characters (such as @,\#), and numerical digits were removed. Finally, the cleansed news article was segmented into sentences. Each sentence was tokenized and each token is labeled with part-of-speech (POS) tags.

\paragraph{\textbf{DKE Extraction}} As mentioned above, DKEs will be used as queries to retrieve candidate papers, so the next step is to extract them from the news article body text. We use DKEs instead of keyphrases or general name entities because DKEs represent the scientific domain knowledge and certain DKEs in news articles also appear the source papers. %Our previous work has demonstrated the efficacy of using DKEs to query a digital library search engine for candidate papers \cite{hoque2019searching}. 
This module is based on a transformer model trained on a multi-domain corpus. We elaborate the DKE extraction in Section~\ref{sec:dkeextraction}.

\paragraph{\textbf{Candidate Paper Retrieval (CPR)}}  This module searches for the scientific paper candidates using extracted DKEs as queries. 
%\hl{[Please check this paragraph carefully-->]}
Here, we assume a frequency-based ranking algorithm such as BM25 \cite{robertson2009probablistic}, which is implemented in popular search platforms, e.g., Apache Solr and Elasticsearch. 
%Even if BM25 is not used in this step, our method will still achieve a comparable performance if the ranker achieves a comparable P@10 (see below). 
In our experiment, we use \url{arXiv.org}, a digital repository that hosts around 2 million pre-printed papers covering major fields in Computer Science, Mathematics, Physics, Astronomy, Statistics, Materials Science, and Social Science. The website offers a free search API. 

To obtain a high recall in this stage, we perform a union of multiple query results. Each query contains a single or a combination of up to 3 DKEs (connected by ``AND''). The final candidate result list is obtained by merging the top 10 results of all queries and removing duplicate papers in terms of titles and authors. This step is necessary for narrowing down the search space because constructing vector representations for \emph{millions of papers} and ranking them by cosine similarity could take impractically long time. This module reduces the candidate pool down to thousands of articles, and boosts the efficiency of the overall retrieval model. Empirically, the system finds approximately 500-3000 candidate papers for each news article. 

\paragraph{\textbf{Paper Re-ranking}}  This module reranks candidate documents based on the vector representations of news articles and paper abstracts. The purpose of this step is to increase the precision by promoting scientific papers that are highly topically relevant to the news article. Here we apply cosine similarity, which is a common practice in many vector search engines (e.g., Covidex \cite{zhang2020covidex}).
%, because we focus on investigating the system performance influenced by document embedding. 
The key is to generate a high-quality vector representation. The vector representation of the scientific paper is constructed by encoding the abstracts into a fix-length vector using a pre-trained language model. The vector representations of the news articles are constructed in a similar way. We investigate the performances of state-of-the-art language models. 

\section{Datasets}
As mentioned above, to our best knowledge, we are not aware of existing datasets containing scientific news and corresponding research papers. Our pilot ground truth dataset is obtained using 100 scientific news articles downloaded from ScienceAlert, ScienceNews, EurekAlert, and Forbes\footnote{ScienceAlert: \url{https://www.sciencealert.com/}; ScienceNews: \url{https://www.sciencenews.org/}; EurekAlert: \url{https://www.eurekalert.org/}; Forbes:\url{www.forbes.com}}. The articles were manually curated so that at least one source scientific paper is provided as a hypertext link or a reference in the original news article. We found up to 5 papers linking to a news article. 
%This dataset doubles our previous effort \cite{hoque2019searching}. 
The average length of these news articles is approximately 900--1000 words. The news articles are from a variety of domains such as history, arts, astronomy, biology, environment, computer science, and medicine. 

We use two datasets for training the DKE extractor: the SemEval~2017 Task~10 dataset \cite{augenstein2017_semeval} and the OA-STM dataset\footnote{\url{https://github.com/elsevierlabs/OA-STM-Corpus}}. Scientific news articles can be written in multiple domains. To train a robust DKE extractor for articles in various domains, we pre-train a model using the SemEval~2017 Task~10 dataset \cite{augenstein2017_semeval}, consisting of 500 passages extracted from journal papers in Computer Science, Materials Science, and Physics. The dataset was double annotated and three types of entities were identified, namely, {\sc material}, {\sc method}, and {\sc process}. There are in total more than 7000 entities manually annotated from the whole dataset. However, this dataset covers only three domains, so we fine-tune the pre-trained model on the OA-STM dataset, containing pre-processed abstracts from scientific papers in 10 domains, including agriculture, astronomy, biology, chemistry, computer science, earth science, engineering, materials science, math, and medicine. There are 11 abstracts in each domain. For each abstract, four core scientific concepts were annotated including {\sc process}, {\sc method}, {\sc material}, and {\sc data}. There are in total 5595 entities annotated. Existing studies indicated that a classifier trained on data from all 10 domains performs better than trained on data from a single domain \cite{brack2020domain}. When using these two datasets, we collapse all categories into one category called DKE. 

\section{Methods}
\subsection{DKE Extraction\label{sec:dkeextraction}}
DKE extraction can be seen as a named entity recognition (NER) task. Although many NER models have been proposed, there is not a consensus that a certain model definitely beats the others in all scenarios. The performance of NER models usually depends on data properties \cite{li2020nersurvey}. Therefore, we compare the following models in the DKE extraction task. 

%\subsubsection{\textbf{Deep Learning-based Models}}
%In our previous work, we compared the performance of multiple sequence tagging models on the task of extracting DKEs from biomedical science papers \cite{wu2020comparative}. Because we change the labeled dataset, we re-evaluated each model and compared their performance on this task. 

\paragraph{\textbf{{BiLSTM-CRF and Res-BiLSTM-CRF}}}
In this model, we applied the bi-directional long short-term memory (BiLSTM) model to obtain the hidden representation of a word level token, followed by a conditional random field (CRF) layer. This model has been applied for many NER tasks \cite{huang2015bidirectional} and achieved state-of-the-art performance on standard datasets, e.g., \cite{luo2020aaai}. We also considered an alternative model with two BiLSTM networks with a residual connection. In the residual unit, the output of a shallow layer is directly added to the output of a deeper layer \cite{srivastava2015highway}. Either model uses random weights as input and learns the hidden representation of each token from the context. 

\paragraph{\textbf{{BiLSTM-W2Vec}}} In this model, the representation of each token is constructed by concatenation of the hidden representation output by a BiLSTM with the pre-trained word2vec model \cite{mikolov2013w2v}. A CRF layer is then applied to classify each token. 

\paragraph{\textbf{{BiLSTM-ChE}}}
Character embedding can be used for capturing morphological information of words \cite{santos2015boosting} and mitigating the out-of-vocabulary problem \cite{verwimp2017character}. In this model, we combine character and word level encodings. The first layer use a BiLSTM to encode each character and combine them into a word-level vector. The second BiLSTM layer encodes each word-level token into a new vector. These two vectors are concatenated to generate the final representation of each work-level token. A CRF classifier is then applied to tag each token.

\paragraph{\textbf{{BiLSTM-ChE-Attention}}}
In this model, a self-attention layer is added after combining the character and word embeddings in the BiLSTM-ChE model. 

\paragraph{\textbf{{Res-BiLSTM-ELMo}}} 
ELMo is a context-dependent language model trained on the 1 Billion Word Benchmark \cite{peters2018deep}, providing word representations with rich features. In this model, we initialize the Res-BiLSTM model using the pre-trained ELMo embedding. 

\begin{figure}
    \centering
    \includegraphics[width=.8\textwidth]{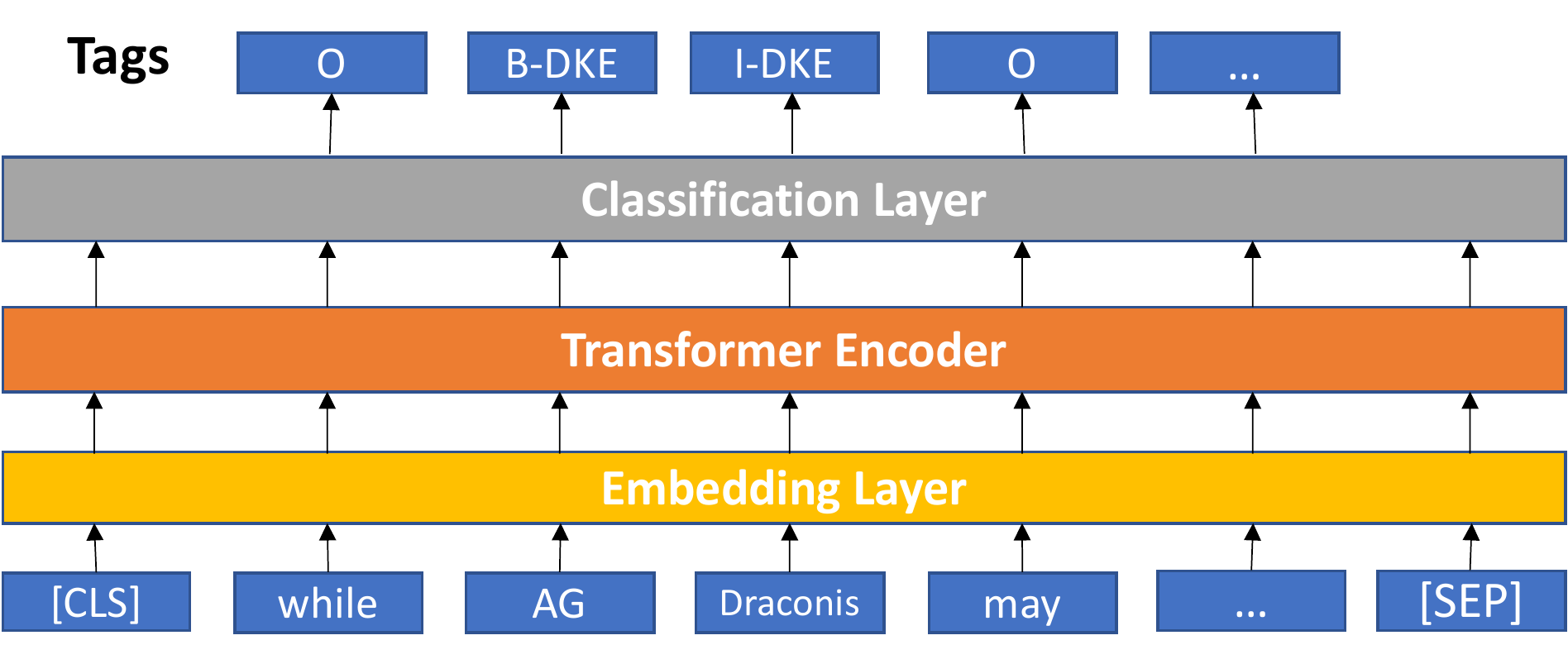}
    \caption{\small Transformer Based DKEs Extractor.}
    \label{fig:transformer_DKE}
\end{figure}

\paragraph{\textbf{Transformer Models}}
The aggregation of self-attention and positional encoding has made transformer models successful for many tasks such as named entity recognition (NER) \cite{vaswani2017attention}. One representative language model is Bidirectional Encoder Representations from Transformers (BERT) \cite{devlin2018bert}, which has been successfully applied for NER \cite{liang2020bond} and other downstream tasks. We implemented two transformer models. One model was trained from scratch on the OA-STM dataset. The other model was developed by fine-tuning a pre-trained BERT model as a backbone encoder on the OA-STM dataset. Before the classification layer, the BERT encoder extracts high-quality language features from our text data. Based on these features, the classification layer classifies these entities into DKEs and non-DKEs (\figlabel~\ref{fig:transformer_DKE}). 

%\subsubsection{\textbf{Baseline Methods}} 
%We also used three baseline models for comparison. These models are unsupervised, machine learning-based, and a model that extracts regular named entities.

\paragraph{\textbf{Text-Rank}} Text-rank is an unsupervised graph-based model inspired by Google's PageRank algorithm to extract keyphrases  \cite{mihalcea2004textrank}. The algorithm builds an undirected graph for each target document, in which the nodes correspond to words in the target document, and edges are drawn between two words that occur next to each other in the text. 

\paragraph{\textbf{HESDK}} HESDK is a hybrid approach to extract DKEs \cite{wu2017hesdk}. In the first phase, candidate phrases are extracted by a grammar-based chunk parser which is then filtered by a linear support vector machine (SVM). In the second phase, a CRF model is used to predict the probabilities of tags for a given token based on lexical and morphological features. The results from both approaches are merged and further filtered by a rule-based filter.

\paragraph{\textbf{Stanford NER}} To demonstrate the advantage of using DKEs, we extract regular named entities using the Stanford CoreNLP  \cite{manning2014stanfordcorenlp}. We use 7-class Stanford NER model trained on the MUC6 and MUC7 datasets. The model extracts seven named entities, including {\sc Location}, {\sc Person}, {\sc Organization}, {\sc Money}, {\sc Percent}, {\sc Date}, and {\sc Time}.

Depending on the length of the news article, the DKE extractors can extract 50-200 DKEs per article, resulting in 500-3000 candidate scientific papers. For all baseline methods, we use the term frequency-inverse document frequency (TFIDF, see below) to represent news articles and scientific papers. 

\subsection{Document Embedding}
In the reranking phase, we represent a news article and scientific paper abstracts with fix-length feature vectors. We compare both local and distributed representation models. First, although pre-trained language models have generally exhibited advantages over local representation models on many tasks, the performance of document representation could be task-dependent. If the similarity of two documents is not on the semantic level but on the literal level, pre-trained language models may loose the advantage. Second, language models on document representation is usually data dependent. The case is more challenging in our task as there is a discrepancy between vocabularies of news articles and research papers. General-use language models, such as BERT \cite{devlin2018bert}, usually well-represent text prevalently used in news and Wikipedia articles. Scientific language models, such as SciBERT \cite{beltagy2019scibert}, on the other hand, usually well-represent text used in scientific papers. 

\paragraph{\textbf{TFIDF weighted Bag-of-Words (BoW)}}
BoW is a traditional text representation model \cite{shahmirzadi2019text}. In this model, each news article or the scientific abstract is represented as a sparse vector containing $|V|$ elements, in which $|V|$ is the vocabulary size of a \emph{retrieval corpus}. Each element is the TFIDF value of the corresponding term. 
The \emph{retrieval corpus} is defined as the combination of the news article and its candidate papers. The IDF for each term is calculated based on the \emph{retrieval corpus} it belongs to. 

\paragraph{\textbf{d2vec}} In this model, for each given document, the vector representation of each word is aggregated in a certain way to represent the whole document. We use the pre-trained word2vec model to calculate a 300 dimensional vector representation of each word. The document vector is the average of vectors of all tokens. 

\paragraph{\textbf{{Doc2vec}}}
Doc2vec is a model to create a vector representation of a document \cite{le2014doc2vec} of various lengths. The \emph{d2vec} model above does not count the word sequence information and does not incorporate the context into the embedded vectors. In doc2vec, when training the word vectors, the document vector $D$ is trained as well. When the document sequence is finished, the document vector $D$ holds a representation of the document. We use the model implemented by Python Gensim Doc2Vec.

\paragraph{\textbf{Weighted Doc2Vec}}
After getting document representation using the Python Gensim Doc2Vec, we extract word representation for each of the words from the document. We then weighted that word using the TFIDF value. Eventually, we combine all the word representation followed by feature-wise averaging to create a new document representation.  

\paragraph{\textbf{SciBERT}}
SciBERT is a transformer-based encoder trained on a large corpus of scientific text \cite{beltagy2019scibert}. Because this model is trained on scientific literature, it has shown advantages over BERT in scientific text classification and summarization tasks \cite{gu2020domain}. partially because its relatively larger vocabulary overlaps with the given corpus. We perform a similar aggregation to the \emph{d2vec} to obtain the document embedding by averaging the vector representation of each token in a document. 

\paragraph{\textbf{SBERT}}
Sentence transformer or sentence-BERT (SBERT) is a modified version of the pre-trained BERT model \cite{reimers2019sentence}. It uses a Siamese network with the triplet loss function to produce sentence embeddings. Each sentence in a document is encoded using SBERT. We then use the averaged embedding as the document embedding. 

\paragraph{\textbf{SPECTER}}
SPECTER is a document embedding model trained on EMNLP scientific publications ranging from 2016 to 2018 \cite{cohan2020specter}. The SPECTER model was designed to overcome the limitations of SciBERT by leveraging inter-document relatedness. This model uses a pre-trained SciBERT transformer model as a backbone and incorporates inter-document context into the SciBERT model. SPECTER builds embeddings from the title and abstract of a paper.

\section{Evaluation and Comparison}
\subsection{Metrics}
We use precision, recall, and F1 score to evaluate the DKEs extractor models. Precision is calculated as the ratio of correctly extracted DKEs divided by the total number of DKEs extracted. The recall is calculated as the ratio of the correctly extracted DKEs divided by the total number of DKEs labeled. F1 score is the harmonic mean of precision and recall.  
We use the following metrics to evaluate the system.
\paragraph{\textbf{Mean-reciprocal-recall (MRR)}} MRR is defined as 
\begin{equation}
{\rm MRR}=\frac{1}{|Q|}\sum_i{\frac{1}{rank(i)}}
\end{equation}
in which $Q$ is the total number of queries, and $rank(i)$ is the rank of a relevant scientific paper. MRR assumes there is only one relevant document in the search results of each query. When evaluating queries corresponding to multiple papers, we use the top-ranked paper to calculate MRR. 

\paragraph{\textbf{Normalized Discounted Cumulative Gain (NDCG)}} We use NDCG with a binary relevance. The metric was used for evaluating cases in which one query returns multiple relevant papers. 

\paragraph{\textbf{P@$K$}} We use the precision at rank $K$ to measure the fraction of relevant scientific papers within certain top results. It can be used when there are multiple relevant papers. We evaluate P@$K$ when $K=1$, $5$, $10$, $20$, and $50$.

\begin{figure}[t]
    \centering
    \includegraphics[width=0.74\textwidth]{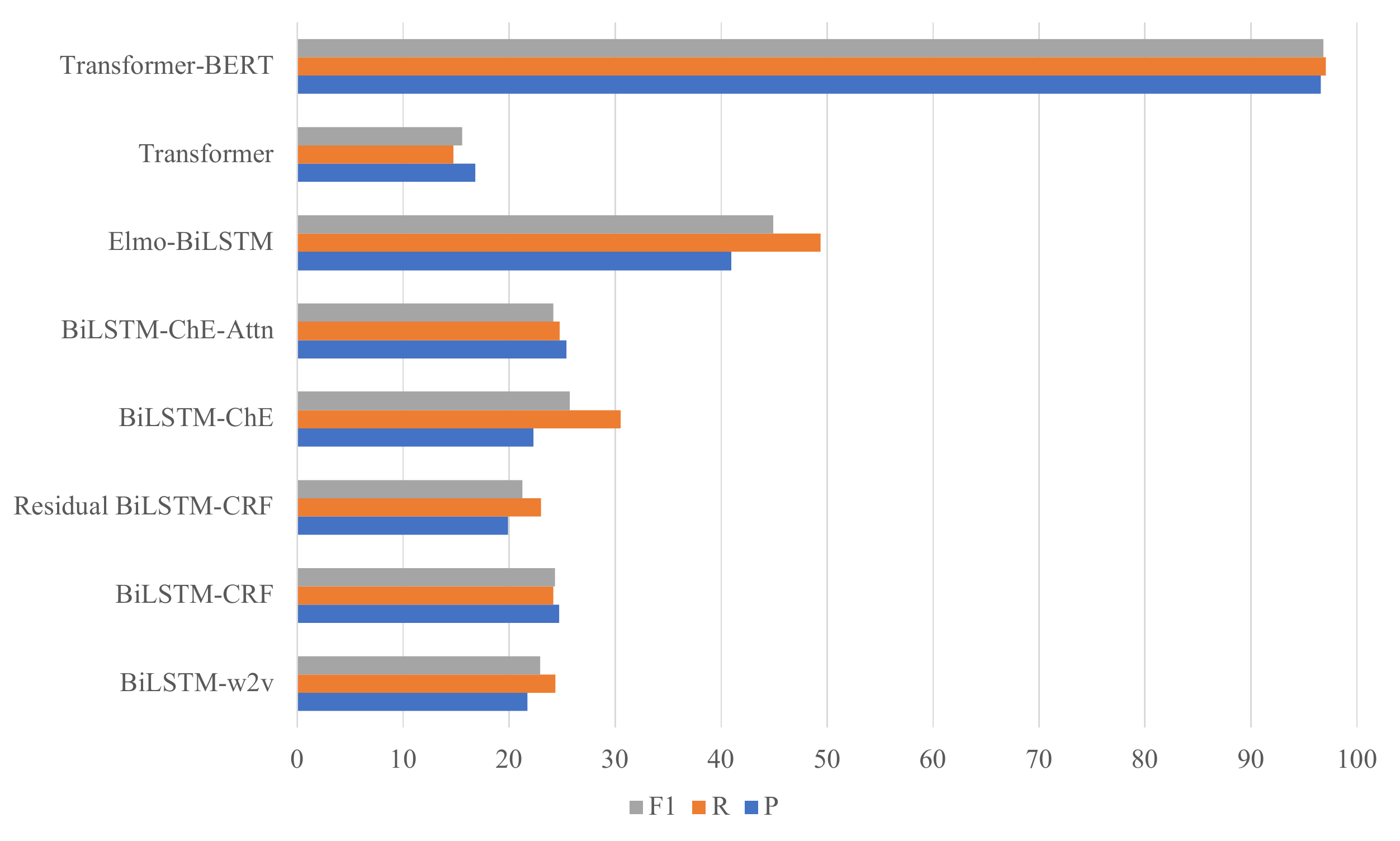}
    \caption{The performance of DKE extraction models.}
    \label{fig:dkeperf}
\end{figure}
\subsection{\textbf{DKE Extraction}} 
DKE extraction is the key step to generate queries to retrieve paper candidates. Each model was trained on 80\% documents from all domains and tested on 20\% of documents on individual domains. \figlabel~\ref{fig:DKEs_Extractor} show the comparison of performance of DKE extraction models. The results indicate that the fine-tuned BERT model outperforms all other models, achieving a nearly perfect performance for all domains, with F1 varying from 0.92 to 1.00 depending on the domain (Table~\ref{fig:DKEs_Extractor}). Specifically, the model correctly extracted all DKEs in the math domain. The superior results are attributed to the BERT transformer encode, which was pre-trained on CoNLL-2003 \cite{devlin2018bert}. 

The other models underperformed most likely because they were trained from scratch on much smaller training datasets and did not generalize well. Among these models, the ELMo-BiLSTM performed relatively better than its peer models. Specifically, the model achieved an F1=52.8\%, 53.5\%, and 51.1\% for materials science, biology, and chemistry, respectively. The results verified the advantage of initializing the BiLSTM encoder with pre-trained language models, e.g., \cite{wu2020comparative}. One interesting phenomenon is that adding self-attention to the BiLSTM-ChE model boosts the performance on certain domains such as agriculture, engineering, math, and biology but decreases the F1 scores of other domains.

\begin{table*}[htbp]
\scriptsize
 \caption{A comparison of our system in different settings.}
\begin{center}
\begin{threeparttable}
\begin{tabular}{c|c|c|c|ccccc|c|c|c|c}
\toprule
     System setting& Query & DKE & Text  & \multicolumn{5}{c|}{P@$K$} & MRR & Average & $T_{\rm PRR}$ & $T_{\rm all}$ \\
     name & type & model& representation & \multicolumn{5}{c|}{$K$=1/5/10/20/50} & & NDCG & (sec) & (sec)\\
\hline

Baseline1\tnote{1} & -- & -- & -- &  14\% & -- & -- & -- & 38\% & -- & -- & -- & --\\
\hline
Baseline2\tnote{2} & KP\tnote{2} & TextRank & TFIDF & {18\%} & {23\%} & {23\%}   & {28\%} & {29\%} & 0.19 & .22 & {\textbf{0.8}} & {21.18} \\
\hline
Baseline3\tnote{3} & NE\tnote{3} & CoreNLP &TFIDF & {38\%} & {44\%}  & {45\%} & {46\%} & {47\%} & 0.40 & 0.44 & {\textbf{0.8}}& {\textbf{13.90}} \\
\hline
Baseline4 & DKE\tnote{4} & HESDK &TFIDF & {39\%}&{45\%}& {49\%} &{55\%}& {60\%} & {0.43} & {0.48}& {\textbf{0.8}} & {116.33}\\
 \hline  
 BERT-TFIDF & DKE & BERT & TFIDF & \textbf{50\%} & \textbf{71\%}  & \textbf{74\%} & {80\%} & {86\%} & \textbf{0.59} & \textbf{0.70} & {\textbf{0.8}} & {66.70} \\
 \hline
  BERT-d2vec & DKE & BERT & d2vec & {20\%} &{37\%}&{41\%}&{55\%}&{69\%}&{0.28}&{0.40}&{4.04}&{ 142.17} \\
 \hline
BERT-Doc2Vec & DKE & BERT& Doc2Vec & {36\%} & {51\%}  & {54\%} & {64\%} & {71\%} & 0.43 & 0.52 & {365.59} & {459.71} \\
\hline
BERT-WDoc2Vec & DKE & BERT& Weighted Doc2Vec & {35\%} & {55\%}  & {60\%} & {68\%} & {84\%} & 0.43 & 0.55 & {350.90}& {473.14}\\
\hline
BERT-SciBERT & DKE & BERT& SciBERT& {19\%} & {30\%}  & {36\%} & {43\%} & {67\%} & 0.25 & 0.36 & {732.74}&{875.88}\\
\hline
BERT-SBERT & DKE & BERT& SBERT & {47\%} & {69\%}  & {74\%} & {82\%} & {90\%} & 0.57 & 0.69 & {10.12} & { 119.43}\\
\hline
BERT-SPECTER & DKE & BERT & SPECTER & {47\%} & {69\%}  & {73\%} & \textbf{84\%} & \textbf{91\%} & 0.57 & 0.69 & {4.22}&{ 110.31} \\
\hline
\end{tabular}
\begin{tablenotes}
 \item[1] Quoted from \cite{harrison2019recommending}. The specific corpus used is not available, making it impossible to make a fair comparison. 
 \item[2] Keyphrases extracted using TextRank \cite{mihalcea2004textrank}.
 \item[3] Named entities extracted using Stanford CoreNLP \cite{manning2014stanfordcorenlp}.
 \item[4] HESDK \cite{wu2017hesdk}.
  \end{tablenotes}
  \end{threeparttable}
\label{tab2}
\end{center}
\end{table*}

\subsubsection{\textbf{System Performance}}
Table~\ref{tab2} shows the performance of the system on retrieving scientific evidence papers. For each system setting, we report the query type, the model used for DKE extraction, and document representation, P@$K$, MRR, and the average NDCG. We also measured the time it takes to finish ranking (the PRR module only), and the overall time for the entire system (\figlabel~\ref{fig:arch}). The runtime information can be important to deploy an online system. For comparison, we add four baseline settings, all using TFIDF weighted BoW model to represent documents but use different query types.

 %Among the baselines, baseline4 which is our previous scholarly paper retrieval model performs better.  
%The BERT-based DKEs extractors along with different document representation methods perform better than any baseline models. 

%The result indicates that the transformer-based retrieval model beats all baselines. 
The results indicate that the BERT-TFIDF and the BERT-SPECTER settings achieved the top performances. The BERT-TFIDF model achieves the best P@1/5/10, MRR, and average NDCG. The BERT-SPECTER model achieves the best P@20 and P@50. 
Specifically, the best baseline (Baseline4), which retrieves 60\% scientific papers within the top 50th position, whereas the BERT-SPECTER model can retrieve 91\%. The result first demonstrates the efficacy of querying DKEs, as opposed to general named entities. In particular, retrieval settings using DKEs as queries outperformed the all retrieval settings using general named entities or keyphrases. Second, the relatively high P@K when K is high (k=20, 50) can be attributed to the powerful capability of the language models to capture the semantic similarities between news articles and papers. On the other hand, the simple TFIDF document presentation coupled with BERT model outperforms all other models when K=1, 5, and 10. This result indicates that the TFIDF model can capture news article and scientific paper pairs that exhibit higher literal similarities. However, when we lower the selection threshold (by increasing $k$), the the most relevant scientific papers to news articles are more semantically similar. We postulate this could be attributed to paraphrasing instead of using exactly the same terms. 

%The reason behind these improvements is the BERT-based DKEs extractor model can extract more meaningful DKEs than our previous DKEs model.

%in terms of the time complexity and performance metrics. Transformer-based document representation models with the same query type also performed well. For certain news articles, the BERT-TFIDF method fails to rank scientific papers within the 50th position, achieving P@$K$=86\%. However, the BERT-SPECTER and SBERT with BERT-based DKE extractors retain more relevant scientific papers (91\%) within the 50th position. From Table~\ref{tab2}, we can see that BERT-SPECTER and BERT-SBERT have higher P@$K$ values when $K=20$ and $50$ than BERT-TFIDF. These results suggest that (1) the transformer-based models can better represent semantics in both the news article and the scientific paper at lower ranks ($K\geq20$), and (2) the TFIDF model is better used for retrieving relevant documents at higher ranks ($K<20$). 
Through error analysis, we found that the major reason that caused our retrieval models to fail was that the scientific news contain much less DKEs. One example is a news article called ``How to spot deepfakes? Look at light reflection in the eyes''\footnote{\url{http://www.buffalo.edu/news/releases/2021/03/010.html}}. Other types of news articles use more images, videos, and equations to convey scientific discoveries, rather than plain text. One example is a news article titled ``Math Genius Has Come Up With a Wildly Simple New Way to Solve Quadratic Equations''\footnote{\url{https://www.sciencealert.com/math-genius-has-come-up-with-a-wildly-simple-new-way-to-solve-quadratic-equations}}.

Regarding runtime, Baseline3 using the Stanford CoreNLP takes the shortest overall time of 13.90 seconds. The top performing setting BERT-TFIDF takes the shortest PRR time of 0.8 second and a relatively short overall time of 66.70 seconds. The BERT-SPECTER setting takes much longer time, which almost doubles the BERT-TFIDF model. From the perspective of building a production service, BERT-TFIDF seems a better choice  but the relatively long runtime is still a bottleneck for building a real-time system.

\section{Conclusion and Future Work} 

%This work extended our previous work on recommending pertinent papers for scientific news articles \cite{hoque2019searching}. 
In this work, we proposed a system called SciEv, which automatically recommends scientific papers given a scientific news article. Although this can be broadly treated as a citation recommender system, we introduced a new scenario, in which the citing document is a news article and the cited document is one or several research papers. Our system consists of four modules: preprocessing, DKE extraction, candidate paper retrieval using DKEs, and paper reranking based on document embedding. We trained a multi-disciplinary transformer-based transfer-learning model that beats other heuristic and learning-based models, achieving an F1=0.93--1. We also compares the capabilities of different document embedding models in capturing the similarities between the news article and research papers. Our experiments on different system settings indicated that using DKEs was an effective and efficient way to retrieve research papers given a scientific news article. However, the TFIDF representation seems more powerful than the language model (e.g., SPECTER) to find our scientific papers when K is relatively low (K$<20$). The language model starts to exhibit advantages over TFIDF when the search results are more inclusive with a higher K (K$\geq20$ in our case). The results indicate that an ensembled re-ranking model may achieve an higher performance, which we will pursue in future work.

The ultimate goal is to build a public application that is capable of automatically assessing the credibility of scientific news, based on pertinent scientific papers. To this end, we need to find effective and efficient ways to find the most relevant ones pertaining to a given scientific news report from the vast amount of scholarly papers and to evaluate the consistency of a scientific report against a list of relevant publications. The SciEv system we proposed here answers the first question. However, the overall runtime is still over 10 seconds. In the future, we will consider a more efficient method to further reduce the overall runtime to seconds by parallelizing queries and text embeddings. 

%%
%% The next two lines define the bibliography style to be used, and
%% the bibliography file.
\bibliographystyle{unsrtnat}
\bibliography{references}

\begin{thebibliography}{44}
\providecommand{\natexlab}[1]{#1}
\providecommand{\url}[1]{\texttt{#1}}
\expandafter\ifx\csname urlstyle\endcsname\relax
  \providecommand{\doi}[1]{doi: #1}\else
  \providecommand{\doi}{doi: \begingroup \urlstyle{rm}\Url}\fi

\bibitem[West and Bergstrom(2021)]{west2021misinformation}
Jevin~D. West and Carl~T. Bergstrom.
\newblock Misinformation in and about science.
\newblock \emph{Proceedings of the National Academy of Sciences}, 118\penalty0
  (15), 2021.
\newblock ISSN 0027-8424.
\newblock \doi{10.1073/pnas.1912444117}.
\newblock URL \url{https://www.pnas.org/content/118/15/e1912444117}.

\bibitem[Zarocostas(2020)]{zarocostas2020infodemic}
John Zarocostas.
\newblock How to fight an infodemic.
\newblock \emph{The Lancet}, 395\penalty0 (10225):\penalty0 676, 2020.
\newblock ISSN 0140-6736.
\newblock \doi{https://doi.org/10.1016/S0140-6736(20)30461-X}.
\newblock URL
  \url{https://www.sciencedirect.com/science/article/pii/S014067362030461X}.

\bibitem[Augenstein(2021)]{augenstein2021determining}
Isabelle Augenstein.
\newblock Determining the credibility of science communication.
\newblock \emph{CoRR}, abs/2105.14473, 2021.
\newblock URL \url{https://arxiv.org/abs/2105.14473}.

\bibitem[F{\"{a}}rber and Jatowt(2020)]{farber2020citation}
Michael F{\"{a}}rber and Adam Jatowt.
\newblock Citation recommendation: approaches and datasets.
\newblock \emph{Int. J. Digit. Libr.}, 21\penalty0 (4):\penalty0 375--405,
  2020.
\newblock \doi{10.1007/s00799-020-00288-2}.
\newblock URL \url{https://doi.org/10.1007/s00799-020-00288-2}.

\bibitem[He et~al.(2010)He, Pei, Kifer, Mitra, and Giles]{he2010context}
Qi~He, Jian Pei, Daniel Kifer, Prasenjit Mitra, and C.~Lee Giles.
\newblock Context-aware citation recommendation.
\newblock In Michael Rappa, Paul Jones, Juliana Freire, and Soumen Chakrabarti,
  editors, \emph{Proceedings of the 19th International Conference on World Wide
  Web, {WWW} 2010, Raleigh, North Carolina, USA, April 26-30, 2010}, pages
  421--430. {ACM}, 2010.
\newblock \doi{10.1145/1772690.1772734}.
\newblock URL \url{https://doi.org/10.1145/1772690.1772734}.

\bibitem[Peng et~al.(2016)Peng, Liu, and Lin]{peng2016news}
Hao Peng, Jing Liu, and Chin{-}Yew Lin.
\newblock News citation recommendation with implicit and explicit semantics.
\newblock In \emph{Proceedings of the 54th Annual Meeting of the Association
  for Computational Linguistics, {ACL} 2016, August 7-12, 2016, Berlin,
  Germany, Volume 1: Long Papers}. The Association for Computer Linguistics,
  2016.
\newblock \doi{10.18653/v1/p16-1037}.
\newblock URL \url{https://doi.org/10.18653/v1/p16-1037}.

\bibitem[Kataria et~al.(2010{\natexlab{a}})Kataria, Mitra, and
  Bhatia]{kataria2010aaai}
Saurabh Kataria, Prasenjit Mitra, and Sumit Bhatia.
\newblock Utilizing context in generative bayesian models for linked corpus.
\newblock In Maria Fox and David Poole, editors, \emph{Proceedings of the
  Twenty-Fourth {AAAI} Conference on Artificial Intelligence, {AAAI} 2010,
  Atlanta, Georgia, USA, July 11-15, 2010}. {AAAI} Press, 2010{\natexlab{a}}.
\newblock URL
  \url{http://www.aaai.org/ocs/index.php/AAAI/AAAI10/paper/view/1883}.

\bibitem[Anderson et~al.(2020)Anderson, Odom, Gray, Jones, Christensen,
  Hollingshead, Hadfield, Evans-Pickett, Frost, Wilson, Davidson, and
  Seeley]{anderson2020case}
P.~Sage Anderson, Aubrey~R. Odom, Hunter~M. Gray, Jordan~B. Jones, William~F.
  Christensen, Todd Hollingshead, Joseph~G. Hadfield, Alyssa Evans-Pickett,
  Megan Frost, Christopher Wilson, Lance~E. Davidson, and Matthew~K. Seeley.
\newblock A case study exploring associations between popular media attention
  of scientific research and scientific citations.
\newblock \emph{PLOS ONE}, 15\penalty0 (7):\penalty0 1--15, 07 2020.
\newblock \doi{10.1371/journal.pone.0234912}.
\newblock URL \url{https://doi.org/10.1371/journal.pone.0234912}.

\bibitem[Hoque et~al.(2019)Hoque, Bradley, Kwan, Chiatti, Li, and
  Wu]{hoque2019searching}
Md~Reshad~Ul Hoque, Dash Bradley, Chiman Kwan, Agnese Chiatti, Jiang Li, and
  Jian Wu.
\newblock Searching for evidence of scientific news in scholarly big data.
\newblock In \emph{Proceedings of the 10th International Conference on
  Knowledge Capture}, pages 251--254, 2019.

\bibitem[He et~al.(2018)He, Fang, Cui, Wu, and Lu]{he2018keyphrase}
Guoxiu He, Junwei Fang, Haoran Cui, Chuan Wu, and Wei Lu.
\newblock Keyphrase extraction based on prior knowledge.
\newblock In \emph{Proceedings of the 18th ACM/IEEE on Joint Conference on
  Digital Libraries}, pages 341--342, 2018.

\bibitem[Bornmann and Mutz(2015)]{bornmann2015growth}
Lutz Bornmann and R{\"u}diger Mutz.
\newblock Growth rates of modern science: A bibliometric analysis based on the
  number of publications and cited references.
\newblock \emph{Journal of the Association for Information Science and
  Technology}, 66\penalty0 (11):\penalty0 2215--2222, 2015.

\bibitem[Melville et~al.(2002)Melville, Mooney, and
  Nagarajan]{melville2002aaai}
Prem Melville, Raymond~J. Mooney, and Ramadass Nagarajan.
\newblock Content-boosted collaborative filtering for improved recommendations.
\newblock In Rina Dechter, Michael~J. Kearns, and Richard~S. Sutton, editors,
  \emph{Proceedings of the Eighteenth National Conference on Artificial
  Intelligence and Fourteenth Conference on Innovative Applications of
  Artificial Intelligence, July 28 - August 1, 2002, Edmonton, Alberta,
  Canada}, pages 187--192. {AAAI} Press / The {MIT} Press, 2002.

\bibitem[Capelle et~al.(2012)Capelle, Frasincar, Moerland, and
  Hogenboom]{capelle2012semantics}
Michel Capelle, Flavius Frasincar, Marnix Moerland, and Frederik Hogenboom.
\newblock Semantics-based news recommendation.
\newblock In \emph{Proceedings of the 2nd international conference on web
  intelligence, mining and semantics}, pages 1--9, 2012.

\bibitem[Kumar et~al.(2017)Kumar, Khattar, Gupta, Gupta, and
  Varma]{kumar2017deep}
Vaibhav Kumar, Dhruv Khattar, Shashank Gupta, Manish Gupta, and Vasudeva Varma.
\newblock Deep neural architecture for news recommendation.
\newblock In \emph{CLEF (Working Notes)}, 2017.

\bibitem[Wang et~al.(2018)Wang, Zhang, Xie, and Guo]{wang2018dkn}
Hongwei Wang, Fuzheng Zhang, Xing Xie, and Minyi Guo.
\newblock Dkn: Deep knowledge-aware network for news recommendation.
\newblock In \emph{Proceedings of the 2018 world wide web conference}, pages
  1835--1844, 2018.

\bibitem[Chu et~al.(2019)Chu, Liu, Sun, and Zhou]{chu2019next}
Qianfeng Chu, Gongshen Liu, Huanrong Sun, and Cheng Zhou.
\newblock Next news recommendation via knowledge-aware sequential model.
\newblock In \emph{China National Conference on Chinese Computational
  Linguistics}, pages 221--232. Springer, 2019.

\bibitem[Huang et~al.(2015{\natexlab{a}})Huang, Wu, Liang, Mitra, and
  Giles]{huang2015neural}
Wenyi Huang, Zhaohui Wu, Chen Liang, Prasenjit Mitra, and C~Giles.
\newblock A neural probabilistic model for context based citation
  recommendation.
\newblock In \emph{Proceedings of the AAAI Conference on Artificial
  Intelligence}, volume~29, 2015{\natexlab{a}}.

\bibitem[Kataria et~al.(2010{\natexlab{b}})Kataria, Mitra, and
  Bhatia]{kataria2010utilizing}
Saurabh Kataria, Prasenjit Mitra, and Sumit Bhatia.
\newblock Utilizing context in generative bayesian models for linked corpus.
\newblock In \emph{Twenty-fourth AAAI conference on artificial intelligence},
  2010{\natexlab{b}}.

\bibitem[Devlin et~al.(2018)Devlin, Chang, Lee, and Toutanova]{devlin2018bert}
Jacob Devlin, Ming-Wei Chang, Kenton Lee, and Kristina Toutanova.
\newblock Bert: Pre-training of deep bidirectional transformers for language
  understanding.
\newblock \emph{arXiv preprint arXiv:1810.04805}, 2018.

\bibitem[Zhang et~al.(2020)Zhang, Gupta, Tang, Han, Pradeep, Lu, Zhang,
  Nogueira, Cho, Fang, and Lin]{zhang2020covidex}
Edwin Zhang, Nikhil Gupta, Raphael Tang, Xiao Han, Ronak Pradeep, Kuang Lu, Yue
  Zhang, Rodrigo Nogueira, Kyunghyun Cho, Hui Fang, and Jimmy Lin.
\newblock Covidex: Neural ranking models and keyword search infrastructure for
  the {COVID-19} open research dataset.
\newblock In \emph{Proceedings of the First Workshop on Scholarly Document
  Processing, SDP@EMNLP 2020, Online, November 19, 2020}, pages 31--41.
  Association for Computational Linguistics, 2020.
\newblock \doi{10.18653/v1/2020.sdp-1.5}.
\newblock URL \url{https://doi.org/10.18653/v1/2020.sdp-1.5}.

\bibitem[Robertson and Zaragoza(2009)]{robertson2009probablistic}
Stephen Robertson and Hugo Zaragoza.
\newblock The probabilistic relevance framework: Bm25 and beyond.
\newblock \emph{Found. Trends Inf. Retr.}, 3\penalty0 (4):\penalty0 333–389,
  April 2009.
\newblock ISSN 1554-0669.
\newblock \doi{10.1561/1500000019}.
\newblock URL \url{https://doi.org/10.1561/1500000019}.

\bibitem[Augenstein et~al.(2017)Augenstein, Das, Riedel, Vikraman, and
  McCallum]{augenstein2017_semeval}
Isabelle Augenstein, Mrinal Das, Sebastian Riedel, Lakshmi Vikraman, and Andrew
  McCallum.
\newblock Semeval 2017 task 10: Scienceie - extracting keyphrases and relations
  from scientific publications.
\newblock In \emph{Proceedings of the 11th International Workshop on Semantic
  Evaluation}, pages 546--555, 2017.

\bibitem[Brack et~al.(2020)Brack, D'Souza, Hoppe, Auer, and
  Ewerth]{brack2020domain}
Arthur Brack, Jennifer D'Souza, Anett Hoppe, S{\"{o}}ren Auer, and Ralph
  Ewerth.
\newblock Domain-independent extraction of scientific concepts from research
  articles.
\newblock In \emph{Advances in Information Retrieval - 42nd European Conference
  on {IR} Research, {ECIR} 2020, Lisbon, Portugal, April 14-17, 2020,
  Proceedings, Part {I}}, volume 12035 of \emph{Lecture Notes in Computer
  Science}, pages 251--266. Springer, 2020.
\newblock \doi{10.1007/978-3-030-45439-5\_17}.
\newblock URL \url{https://doi.org/10.1007/978-3-030-45439-5\_17}.

\bibitem[Li et~al.(2022)Li, Sun, Han, and Li]{li2020nersurvey}
Jing Li, Aixin Sun, Jianglei Han, and Chenliang Li.
\newblock A survey on deep learning for named entity recognition.
\newblock \emph{IEEE Transactions on Knowledge and Data Engineering},
  34\penalty0 (1):\penalty0 50--70, 2022.
\newblock \doi{10.1109/TKDE.2020.2981314}.

\bibitem[Huang et~al.(2015{\natexlab{b}})Huang, Xu, and
  Yu]{huang2015bidirectional}
Zhiheng Huang, Wei Xu, and Kai Yu.
\newblock Bidirectional lstm-crf models for sequence tagging.
\newblock \emph{arXiv preprint arXiv:1508.01991}, 2015{\natexlab{b}}.

\bibitem[Luo et~al.(2020)Luo, Xiao, and Zhao]{luo2020aaai}
Ying Luo, Fengshun Xiao, and Hai Zhao.
\newblock Hierarchical contextualized representation for named entity
  recognition.
\newblock In \emph{The Thirty-Fourth {AAAI} Conference on Artificial
  Intelligence, {AAAI} 2020, The Thirty-Second Innovative Applications of
  Artificial Intelligence Conference, {IAAI} 2020, The Tenth {AAAI} Symposium
  on Educational Advances in Artificial Intelligence, {EAAI} 2020, New York,
  NY, USA, February 7-12, 2020}, pages 8441--8448. {AAAI} Press, 2020.
\newblock URL \url{https://aaai.org/ojs/index.php/AAAI/article/view/6363}.

\bibitem[Srivastava et~al.(2015)Srivastava, Greff, and
  Schmidhuber]{srivastava2015highway}
Rupesh~Kumar Srivastava, Klaus Greff, and J{\"{u}}rgen Schmidhuber.
\newblock Highway networks.
\newblock \emph{CoRR}, abs/1505.00387, 2015.
\newblock URL \url{http://arxiv.org/abs/1505.00387}.

\bibitem[Mikolov et~al.(2013)Mikolov, Yih, and Zweig]{mikolov2013w2v}
Tomas Mikolov, Wen{-}tau Yih, and Geoffrey Zweig.
\newblock Linguistic regularities in continuous space word representations.
\newblock In \emph{Processings of 2013 Human Language Technologies: Conference
  of the North American Chapter of the Association of Computational
  Linguistics}, pages 746--751, 2013.

\bibitem[Santos and Guimaraes(2015)]{santos2015boosting}
Cicero Nogueira~dos Santos and Victor Guimaraes.
\newblock Boosting named entity recognition with neural character embeddings.
\newblock \emph{arXiv preprint arXiv:1505.05008}, 2015.

\bibitem[Verwimp et~al.(2017)Verwimp, Pelemans, Wambacq,
  et~al.]{verwimp2017character}
Lyan Verwimp, Joris Pelemans, Patrick Wambacq, et~al.
\newblock Character-word lstm language models.
\newblock \emph{arXiv preprint arXiv:1704.02813}, 2017.

\bibitem[Peters et~al.(2018)Peters, Neumann, Iyyer, Gardner, Clark, Lee, and
  Zettlemoyer]{peters2018deep}
Matthew~E Peters, Mark Neumann, Mohit Iyyer, Matt Gardner, Christopher Clark,
  Kenton Lee, and Luke Zettlemoyer.
\newblock Deep contextualized word representations.
\newblock \emph{arXiv preprint arXiv:1802.05365}, 2018.

\bibitem[Vaswani et~al.(2017)Vaswani, Shazeer, Parmar, Uszkoreit, Jones, Gomez,
  Kaiser, and Polosukhin]{vaswani2017attention}
Ashish Vaswani, Noam Shazeer, Niki Parmar, Jakob Uszkoreit, Llion Jones,
  Aidan~N Gomez, Lukasz Kaiser, and Illia Polosukhin.
\newblock Attention is all you need.
\newblock \emph{arXiv preprint arXiv:1706.03762}, 2017.

\bibitem[Liang et~al.(2020)Liang, Yu, Jiang, Er, Wang, Zhao, and
  Zhang]{liang2020bond}
Chen Liang, Yue Yu, Haoming Jiang, Siawpeng Er, Ruijia Wang, Tuo Zhao, and Chao
  Zhang.
\newblock Bond: Bert-assisted open-domain named entity recognition with distant
  supervision.
\newblock In \emph{Proceedings of the 26th ACM SIGKDD International Conference
  on Knowledge Discovery \& Data Mining}, pages 1054--1064, 2020.

\bibitem[Mihalcea and Tarau(2004)]{mihalcea2004textrank}
Rada Mihalcea and Paul Tarau.
\newblock Textrank: Bringing order into text.
\newblock In \emph{Proceedings of the 2004 Conference on Empirical Methods in
  Natural Language Processing , {EMNLP} 2004}, pages 404--411, 2004.

\bibitem[Wu et~al.(2017)Wu, Choudhury, Chiatti, Liang, and Giles]{wu2017hesdk}
Jian Wu, Sagnik~Ray Choudhury, Agnese Chiatti, Chen Liang, and C~Lee Giles.
\newblock Hesdk: A hybrid approach to extracting scientific domain knowledge
  entities.
\newblock In \emph{Proceedings of the 17th JCDL}, pages 241--244, 2017.

\bibitem[Manning et~al.(2014)Manning, Surdeanu, Bauer, Finkel, Bethard, and
  McClosky]{manning2014stanfordcorenlp}
Christopher~D. Manning, Mihai Surdeanu, John Bauer, Jenny Finkel, Steven~J.
  Bethard, and David McClosky.
\newblock The {Stanford} {CoreNLP} natural language processing toolkit.
\newblock In \emph{Association for Computational Linguistics (ACL) System
  Demonstrations}, pages 55--60, 2014.
\newblock URL \url{http://www.aclweb.org/anthology/P/P14/P14-5010}.

\bibitem[Beltagy et~al.(2019)Beltagy, Lo, and Cohan]{beltagy2019scibert}
Iz~Beltagy, Kyle Lo, and Arman Cohan.
\newblock Scibert: A pretrained language model for scientific text.
\newblock \emph{arXiv preprint arXiv:1903.10676}, 2019.

\bibitem[Shahmirzadi et~al.(2019)Shahmirzadi, Lugowski, and
  Younge]{shahmirzadi2019text}
Omid Shahmirzadi, Adam Lugowski, and Kenneth Younge.
\newblock Text similarity in vector space models: a comparative study.
\newblock In \emph{2019 18th IEEE International Conference On Machine Learning
  And Applications (ICMLA)}, pages 659--666. IEEE, 2019.

\bibitem[Le and Mikolov(2014)]{le2014doc2vec}
Quoc~V. Le and Tom{\'{a}}s Mikolov.
\newblock Distributed representations of sentences and documents.
\newblock In \emph{Proceedings of the 31th International Conference on Machine
  Learning, {ICML} 2014, Beijing, China, 21-26 June 2014}, volume~32 of
  \emph{{JMLR} Workshop and Conference Proceedings}, pages 1188--1196.
  JMLR.org, 2014.
\newblock URL \url{http://proceedings.mlr.press/v32/le14.html}.

\bibitem[Gu et~al.(2020)Gu, Tinn, Cheng, Lucas, Usuyama, Liu, Naumann, Gao, and
  Poon]{gu2020domain}
Yu~Gu, Robert Tinn, Hao Cheng, Michael Lucas, Naoto Usuyama, Xiaodong Liu,
  Tristan Naumann, Jianfeng Gao, and Hoifung Poon.
\newblock Domain-specific language model pretraining for biomedical natural
  language processing.
\newblock \emph{CoRR}, abs/2007.15779, 2020.
\newblock URL \url{https://arxiv.org/abs/2007.15779}.

\bibitem[Reimers and Gurevych(2019)]{reimers2019sentence}
Nils Reimers and Iryna Gurevych.
\newblock Sentence-bert: Sentence embeddings using siamese bert-networks.
\newblock \emph{arXiv preprint arXiv:1908.10084}, 2019.

\bibitem[Cohan et~al.(2020)Cohan, Feldman, Beltagy, Downey, and
  Weld]{cohan2020specter}
Arman Cohan, Sergey Feldman, Iz~Beltagy, Doug Downey, and Daniel~S Weld.
\newblock Specter: Document-level representation learning using
  citation-informed transformers.
\newblock In \emph{Proceedings of the 58th Annual Meeting of the Association
  for Computational Linguistics}, pages 2270--2282, 2020.

\bibitem[Wu et~al.(2020)Wu, Hoque, Reiske, Weigle, Bradshaw, Gaff, Li, and
  Kwan]{wu2020comparative}
Jian Wu, Md~Reshad~Ul Hoque, Gunnar~W Reiske, Michele~C Weigle, Brenda~T
  Bradshaw, Holly~D Gaff, Jiang Li, and Chiman Kwan.
\newblock A comparative study of sequential tagging methods for domain
  knowledge entity recognition in biomedical papers.
\newblock 2020.

\bibitem[Harrison et~al.(2019)Harrison, Martin, Surian, and
  Dunn]{harrison2019recommending}
Eliza Harrison, Paige Martin, Didi Surian, and Adam~G Dunn.
\newblock Recommending research articles to consumers of online vaccination
  information.
\newblock \emph{arXiv preprint arXiv:1904.11886}, 2019.

\end{thebibliography}

%%
%% If your work has an appendix, this is the place to put it.
\appendix

\section{Appendix: DKE Extraction Performance Comparison}
\begin{figure*}
\centering
\begin{tabular}{cc}
  \includegraphics[width=0.45\linewidth, height=5.5cm]{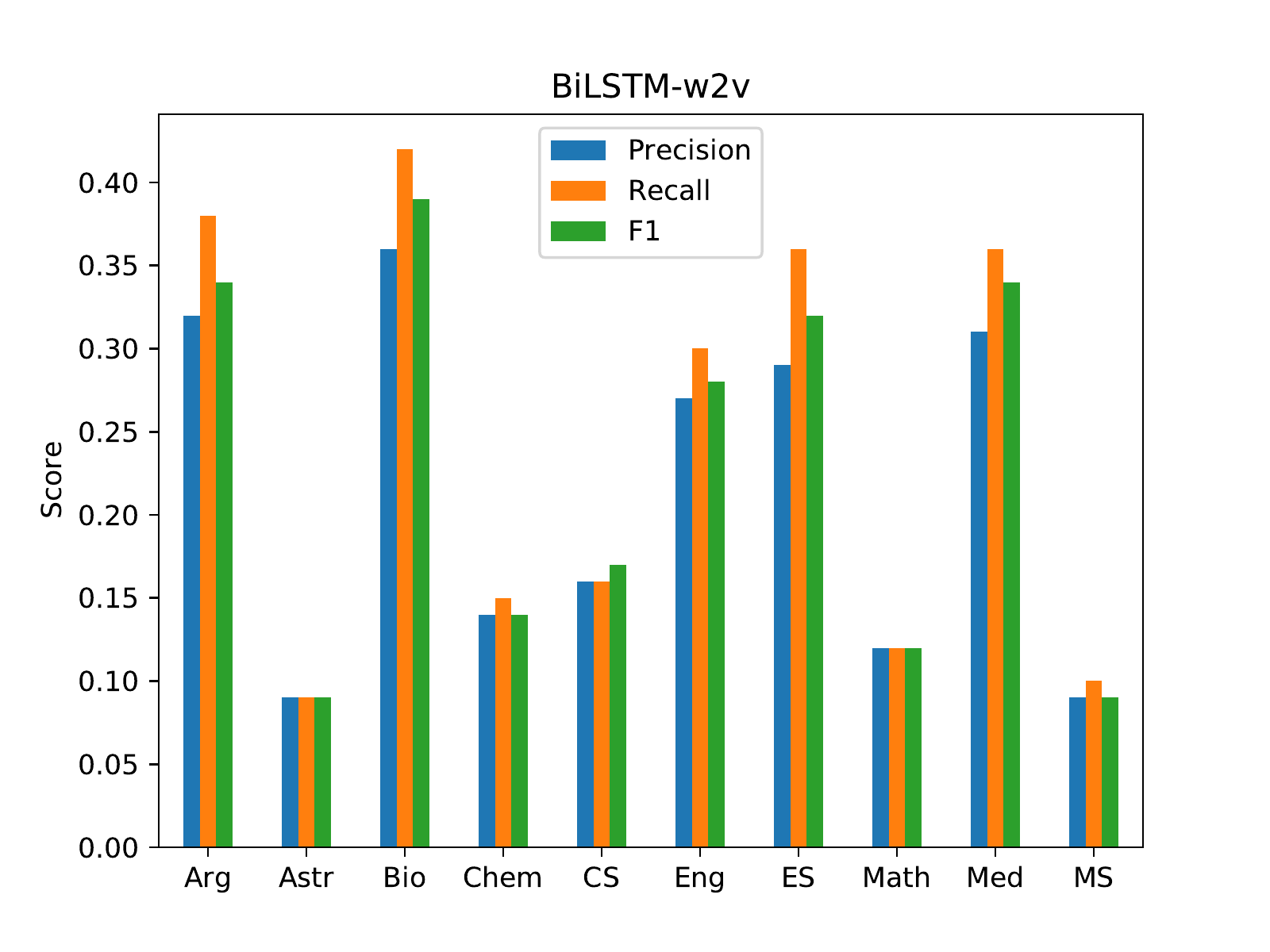} &
  \includegraphics[width=0.45\linewidth, height=5.5cm]{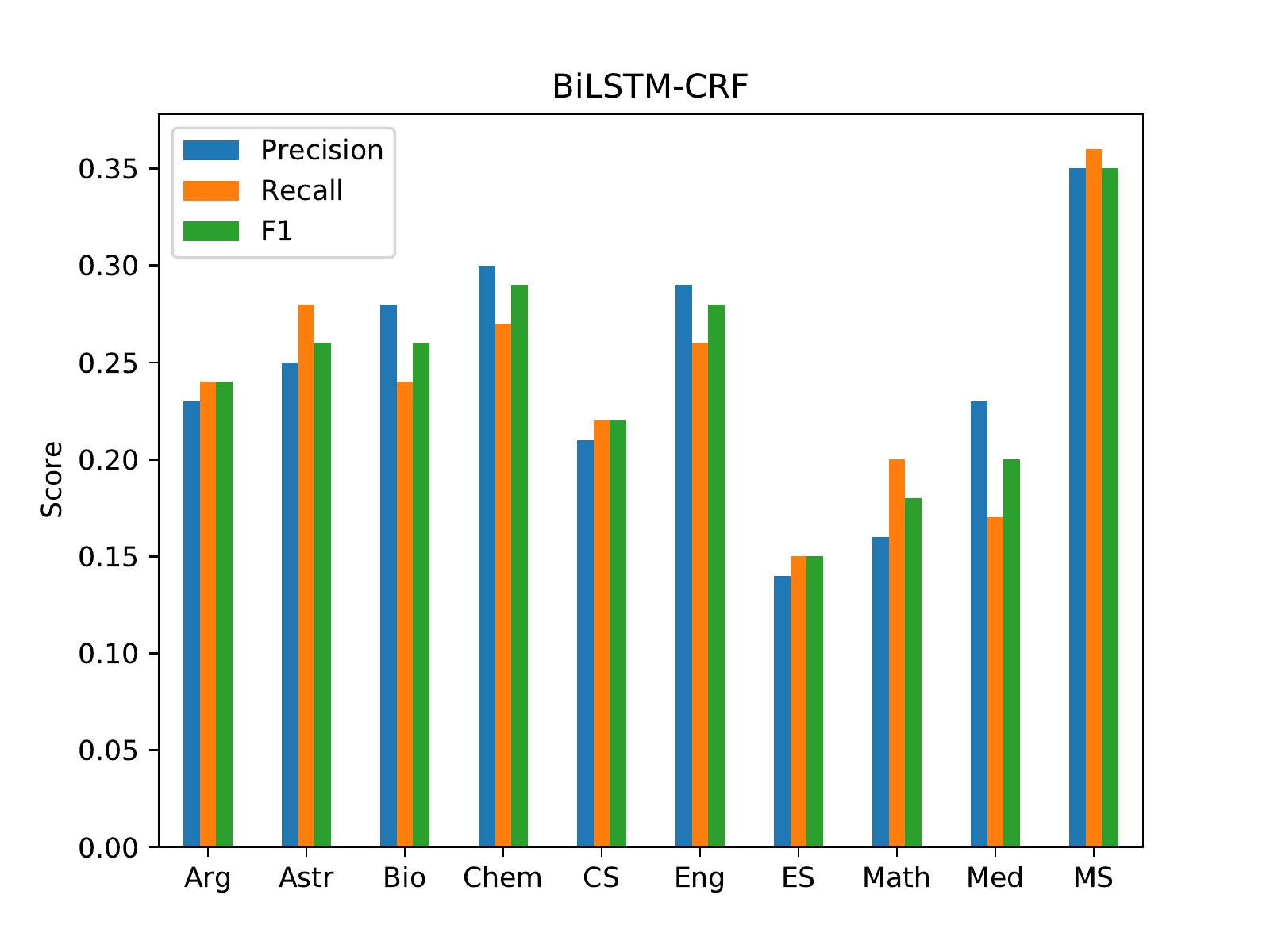} \\
  %(a) BiLSTM-w2v  & (b) BiLSTM-CRF \\
  \includegraphics[width=0.45\linewidth, height=5.5cm]{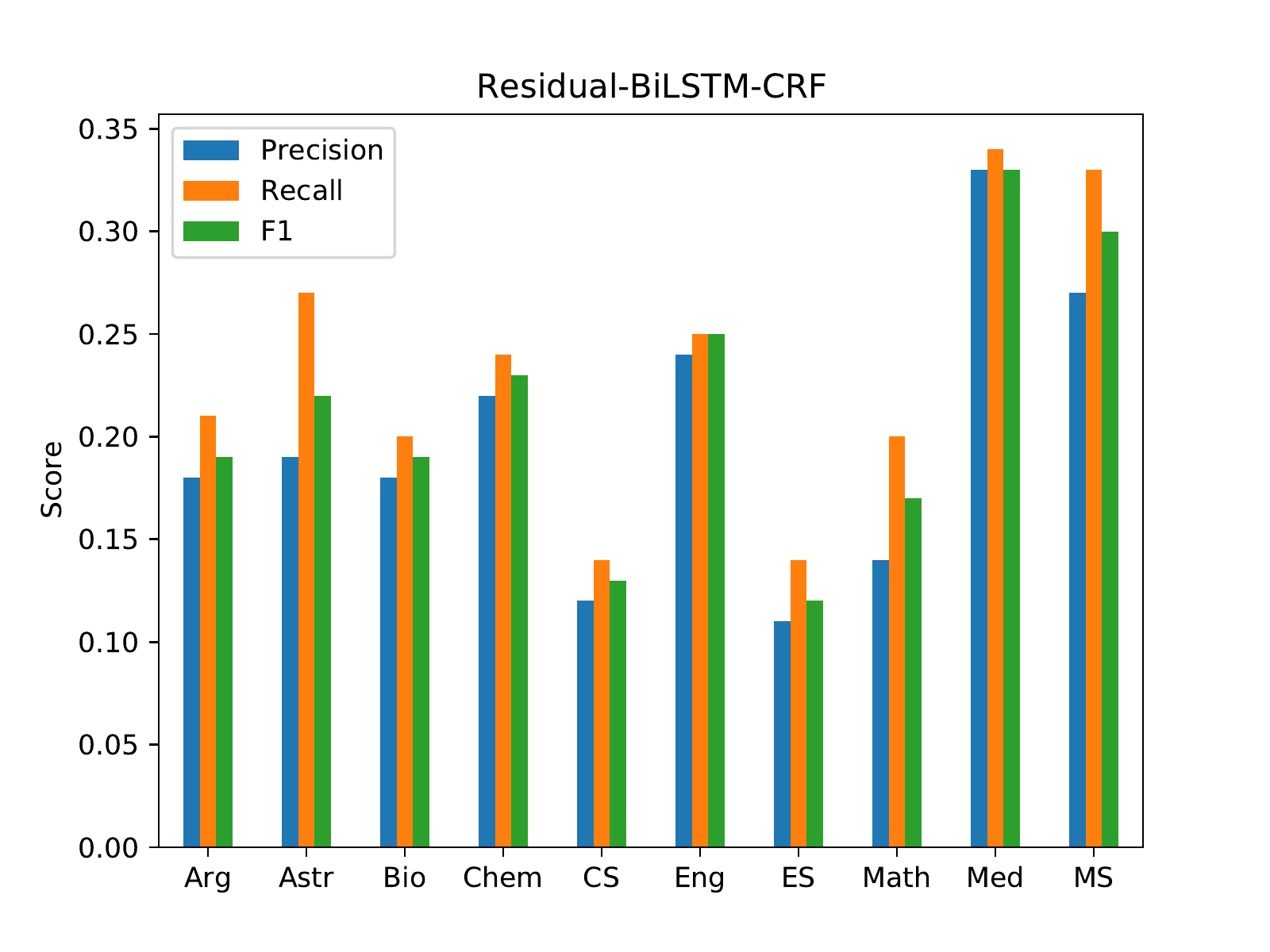} &
  \includegraphics[width=0.45\linewidth, height=5.5cm]{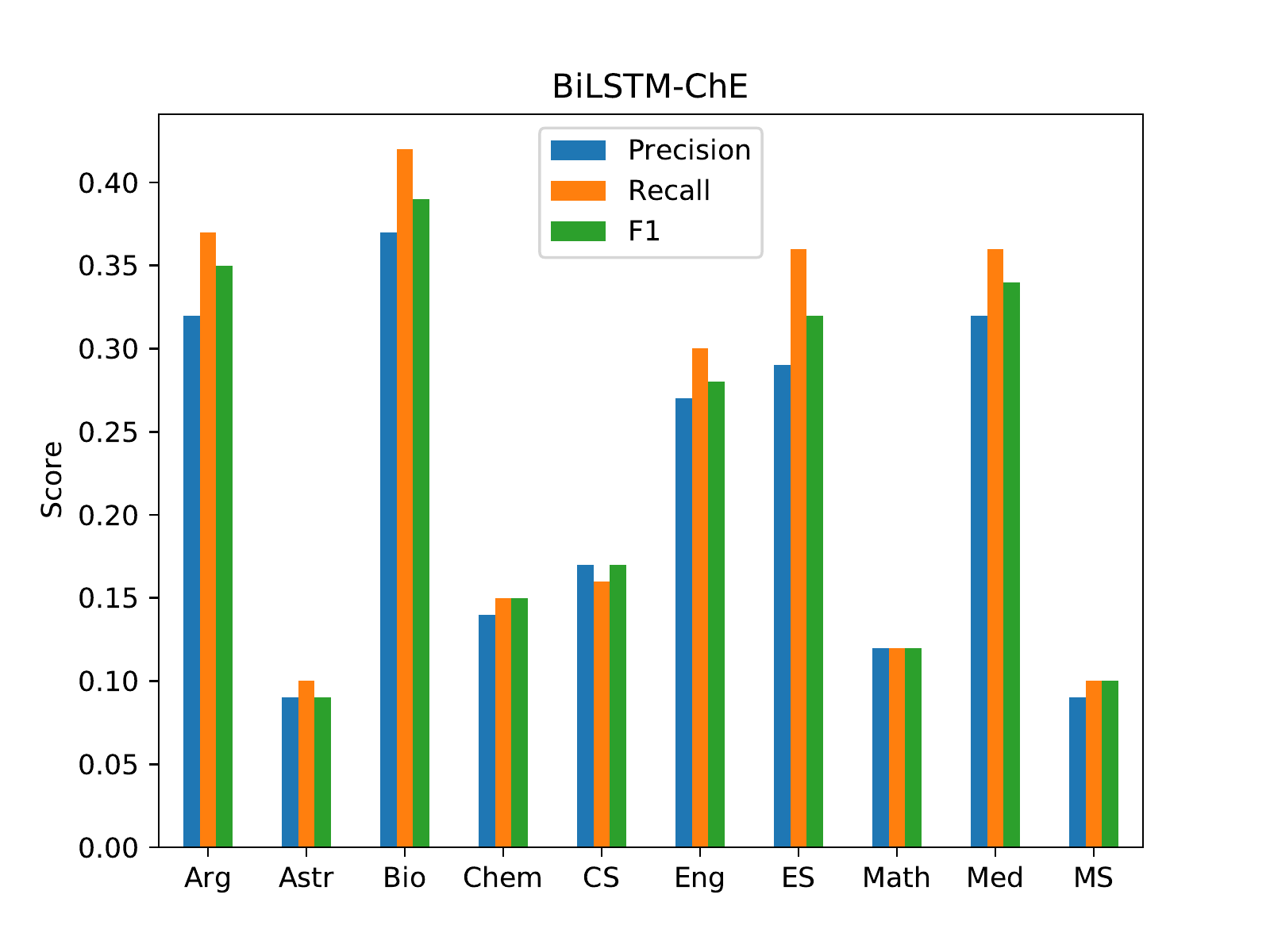} \\
  %(c) Residual-BiLSTM-CRF  & (d) BiLSTM-ChE \\
  \includegraphics[width=0.45\linewidth, height=5.5cm]{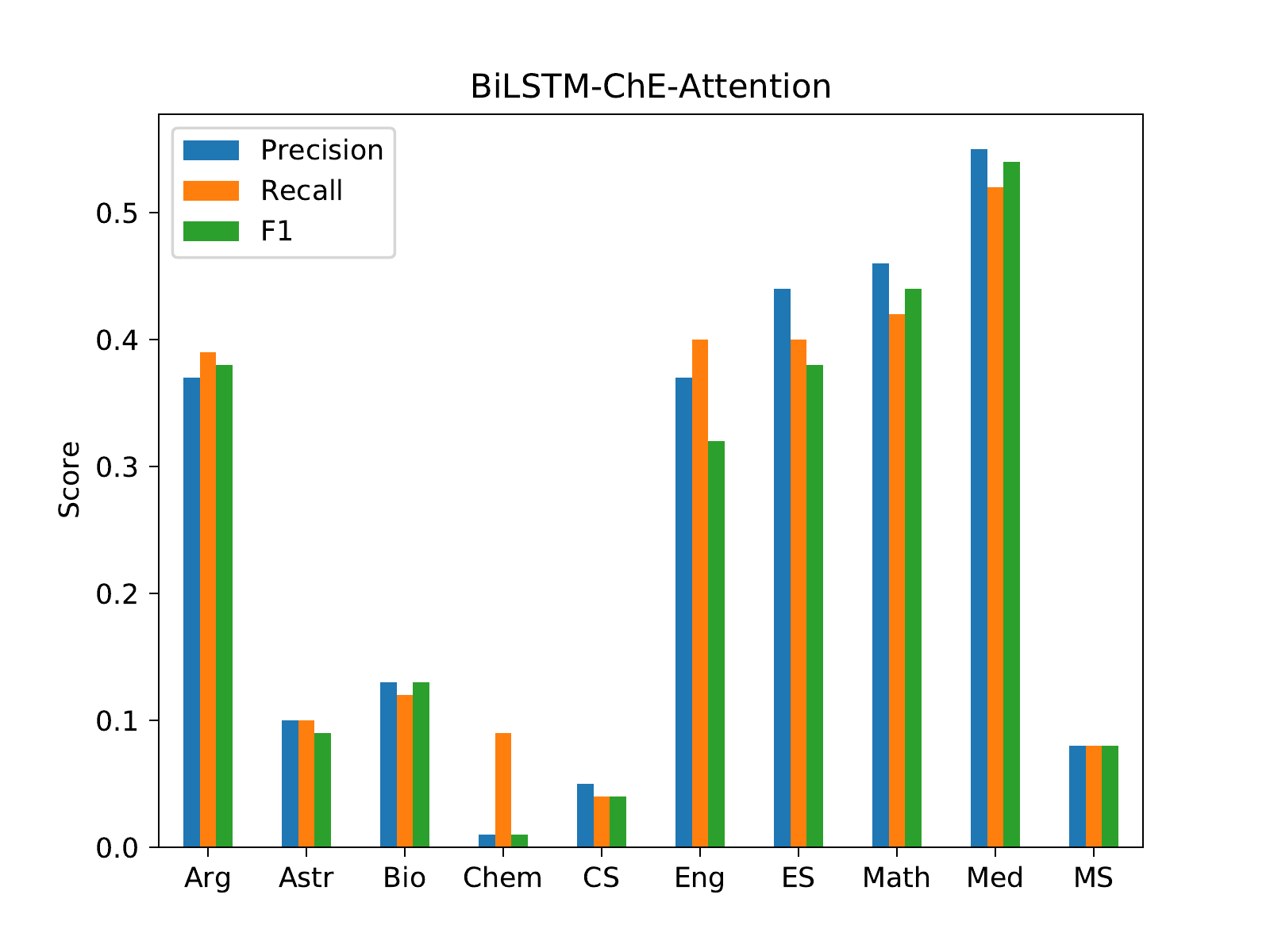} &
  \includegraphics[width=0.45\linewidth, height=5.5cm]{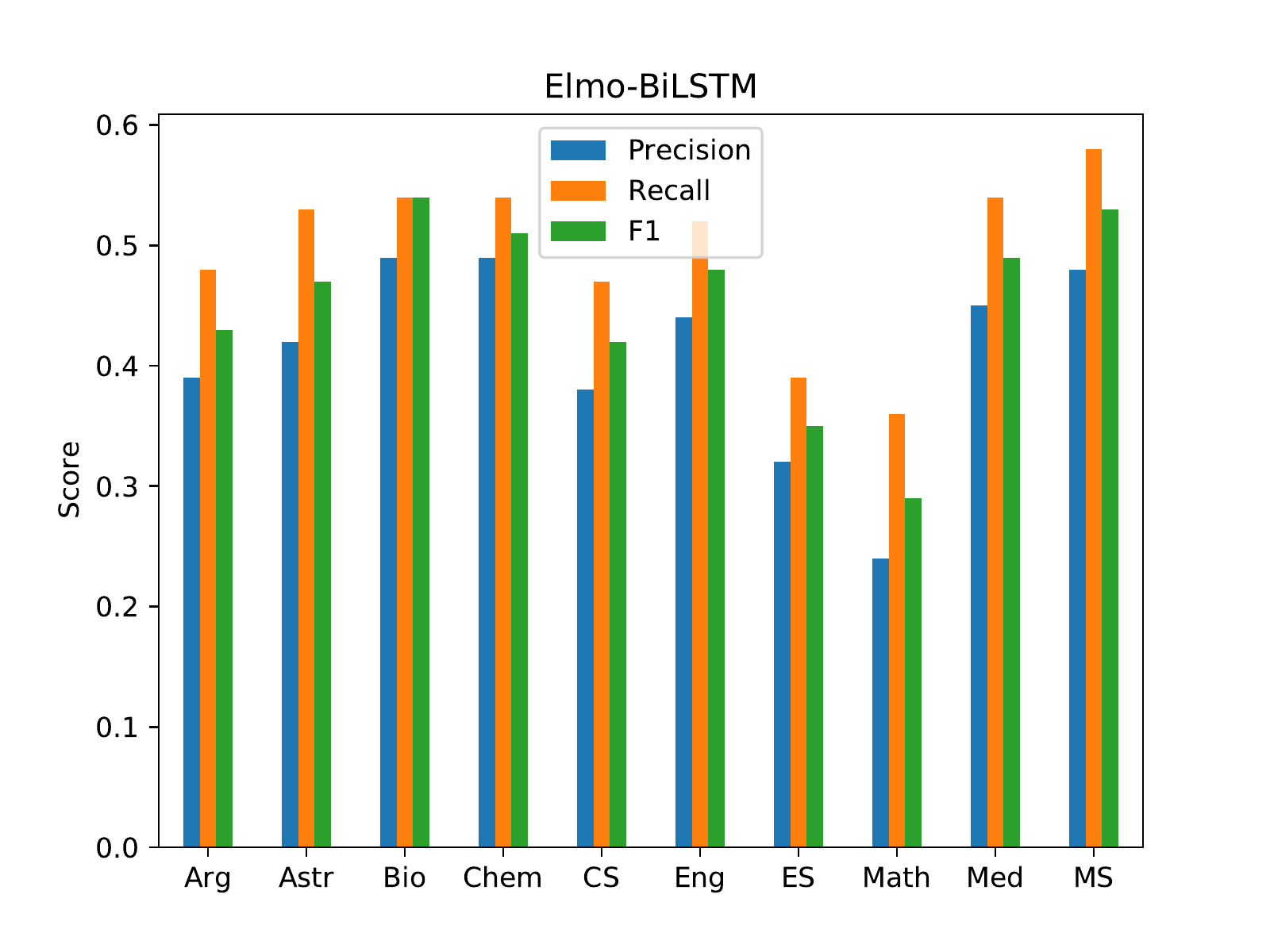} \\
  %(e) BiLSTM-ChE-Attention  & (f) Elmo-BiLSTM \\
  \includegraphics[width=0.45\linewidth, height=5.5cm]{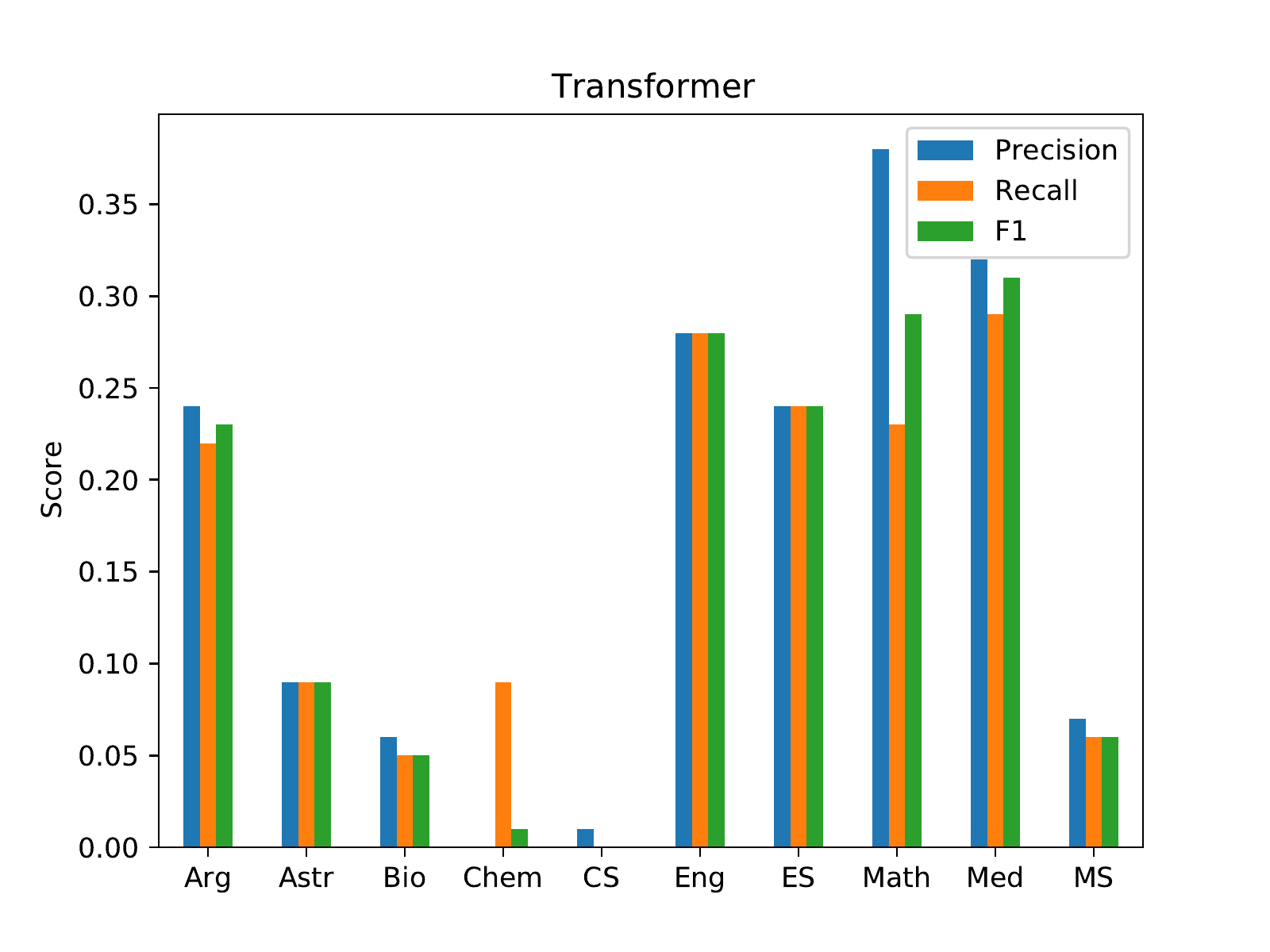} &
  \includegraphics[width=0.45\linewidth, height=5.5cm]{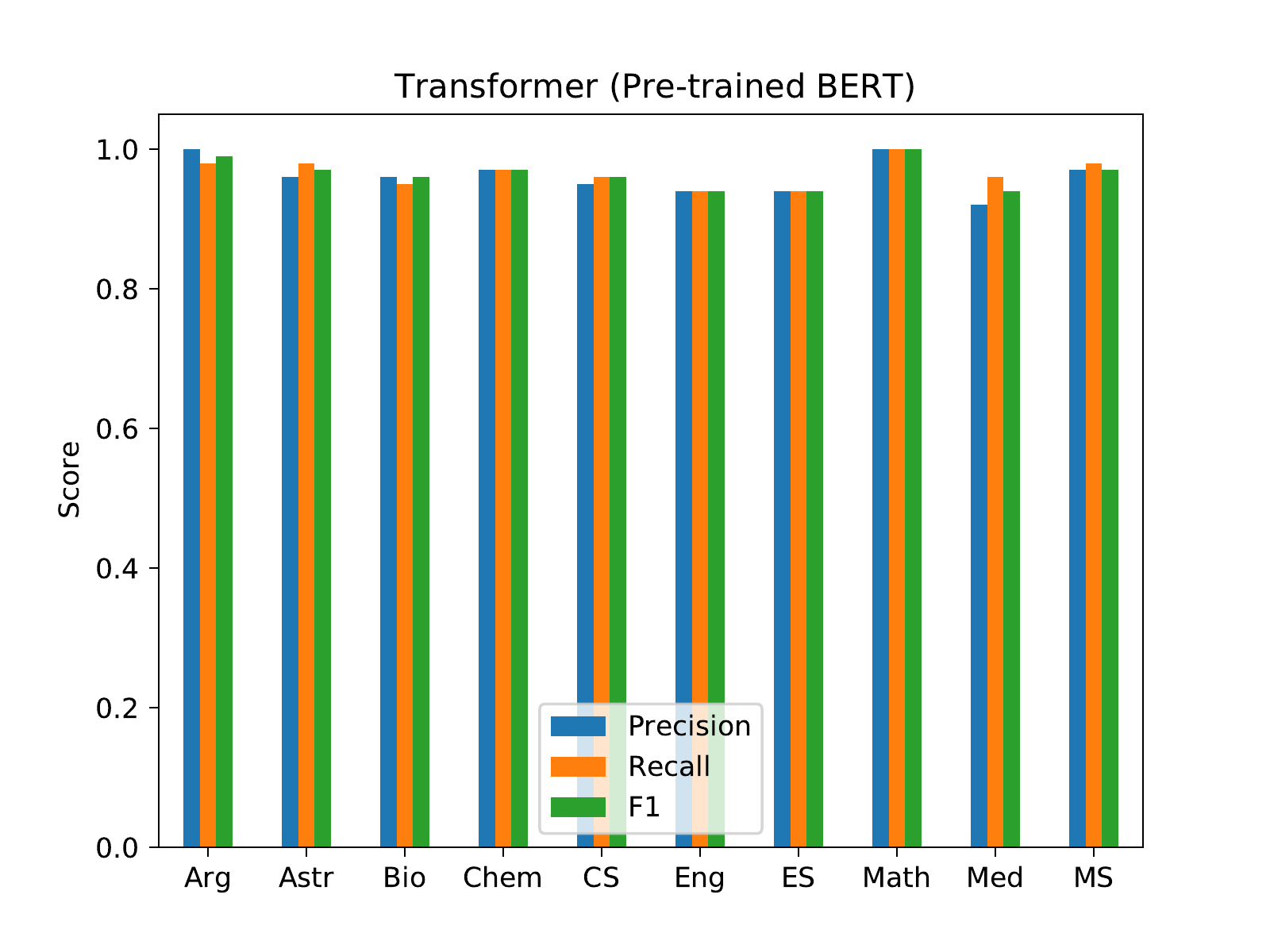} \\
  %(g) Transformer  & (h) Transformer (Pre-trained BERT) \\
\end{tabular}
\caption{A comparison of performances of DKE extraction models. For comparison, we also show a ``Transformer'' model without initializing tokens using the pre-trained BERT model. Categories along x-axis are below. Arg: agriculture; Astr: astronomy; Bio: biology; Chem: chemistry; CS: computer science; Eng: engineering; ES: environmental science; Math: mathematics; Med: medical science; MS: materials science.}
\label{fig:DKEs_Extractor}
\end{figure*}

Figure~\ref{fig:DKEs_Extractor} shows the differential performance of the DKE extractor for 10 domains in the OA-STM dataset. 

\end{document}